\documentstyle[12pt]{article}

\newcommand{\ccr}[2]{{[} {#1},{#2} {]} }        %Kommutator
\newcommand{\CMP}[1]{{\em Commun. Math. Phys.} {\bf {#1}}}
\newcommand{\JMP}[1]{{\em J.~Math. Phys.} {\bf {#1}}}
\newcommand{\NP}[1]{{\em Nucl.~Phys.~B} {\bf {#1}}}

\newcommand{\PL}[1]{{\em Phys. Lett.} {\bf {#1}}}

\newcommand{\eq}{\begin{equation}}
\newcommand{\eqend}{\end{equation}}
\newcommand{\eqa}{\begin{eqnarray}}
\newcommand{\neqa}{\begin{eqnarray*}}
\newcommand{\eqaend}{\end{eqnarray}}
\newcommand{\neqaend}{\end{eqnarray*}}

\newcommand{\bma}[1]{\begin{array}{#1}}
\newcommand{\ema}{\end{array}}
\newcommand{\bc}{\begin{center}}
\newcommand{\ec}{\end{center}}

\newcommand{\HH}{{\rm H}}

\newcommand{\U}{{\rm U}}
\newcommand{\SU}{{\rm SU}}
\newcommand{\Om}{\Omega}
\newcommand{\del}{\delta}
\newcommand{\cF}{{\cal F}}
\newcommand{\normal}[1]{:{#1}:}
\newcommand{\Ref}[1]{(\ref{#1})}
\newcommand{\dd}{\mbox{${\rm d}$}}
\newcommand{\ee}[1]{\mbox{{\rm e}}^{#1}}
\newcommand{\ii}{{\rm i}}
\newcommand{\R}{{\sf I} \! {\sf R}}

\newcommand{\Z}{{\sf Z} \! \! {\sf Z}}
\newcommand{\f}{\frac}
\newcommand{\cH}{{\cal H}}
\newcommand{\cG}{{\cal G}}
\newcommand{\cA}{{\cal A}}
\newcommand{\tra}[1]{{\rm tr} ({#1})}

\begin{document}
\pagestyle{empty}
%\begin{flushright}
%{UBCTP 92-\\
%November 1992}
%\end{flushright}
%%\vspace{.4cm}
\renewcommand{\thefootnote}{\alph{footnote}}
\begin{center}

{\Large \bf Gribov ambiguity and non-trivial vacuum structure of gauge
theories on a cylinder}\\
\vspace{1 cm}

{\large Edwin Langmann}\footnote{supported by the ``Fonds zur
F\"orderung der wissenschaftlichen Forschung'' under the contract Nr.\
J0789-PHY}\\ {\large and Gordon W.  Semenoff}\footnote{supported in
part by the Natural Sciences and Engineering Research Council of
Canada.} \\
\vspace{0.3 cm}
{\em Department of Physics, The University of British Columbia\\
Vancouver, B.C., V6T 1Z1, Canada }\\

\end{center}
%%\vspace{0.4cm} \noindent

\setcounter{footnote}{0}
\renewcommand{\thefootnote}{\arabic{footnote}}

\begin{abstract}
Using the hamiltonian framework, we analyze the Gribov problem for
$\U(N)$ and $\SU(N)$ gauge theories on a cylinder (= (1+1) dimensional
spacetime with compact space $S^1$).  The space of gauge orbits is
found to be an orbifold. We show by explicit construction
that a proper treatment of the Gribov ambiguity leads to a highly
non-trivial structure of all physical states in these quantum field
theory models. The especially interesting example of massless QCD is
discussed in more detail: There, some of the special static gauge
transformations which are responsible for the Gribov ambiguity also
lead to a spectral flow, and this implies a chiral condensate in all
physical states. We also show that the latter is closely related to
the Schwinger term and the chiral anomaly.

\end{abstract}

\newpage
\pagestyle{plain}
\setcounter{page}{1}

%%%%%%%%%%%%%%%%%%%%%%%%%%%%%%%%%%%%%%%%%%%%%%%%%%%%

{}{\bf 1.} The elimination of gauge degrees of freedom is essential
for understanding and extracting physical information from Yang-Mills
(YM) gauge theories. This is usually done by `fixing a gauge', i.e.\
requiring that the field configurations obey some gauge condition such
as the Coulomb or the Landau gauge (e.g.\ in the path integral
formalism by means of the Faddeev-Popov trick). Though adequate for
perturbative calculations, it is well-known that such a procedure is
not sufficient for a deeper understanding of non-Abelian YM theories:
It has been pointed out already by Gribov \cite{G} that there are many
gauge equivalent configurations obeying the Coulomb or the Landau
gauge condition, and that the existence of these should play a crucial
role for non-perturbative features of these theories such as
confinement.  Later on it was shown by Singer \cite{S} that in 4
(compact) spacetime dimensions such a Gribov ambiguity arises for any
reasonable gauge fixing condition. It has also been found that the
lack of a full understanding of Gribov ambiguities and their proper
treatment is one major obstacle to a rigorous non-perturbative
construction of gauge theories in 4 dimensions \cite{Riv}. (For a
recent discussion of the Gribov problem see \cite{B}.)

In 2 spacetime dimensions gauge theories are much simpler.  Pure YM
theory on a plane $\R\times\R$ is trivial with no propagating degrees
of freedom.  However, on a manifold with nontrivial topology such as a
Riemann surface \cite{W,BT} or (in the hamiltonain apprach) a cylinder
(spacetime $S^1\times\R$)
\cite{Rajeev,HH,M,LS}, it has a finite number of physical degrees
of freedom and several non-trivial features which makes it an
interesting toy model.  Moreover, YM gauge theories with matter
(fermions or bosons) on a cylinder allow for a (essentially) rigorous
construction on an operator level by means of the theory of quasi-free
representations of fermion and boson field algebras \cite{CR,GL} (we
are planning to report on that in more detail in future publications;
for an alternative approach see \cite{M}).  This corresponds to the
well-known fact that on a cylinder, normal ordering of the bilinears
of the matter fields with respect to the {\em free} vacuum is
sufficient to eliminate all divergences in the fully interacting gauge
theory.

In this paper we shall examine the Gribov problem for $\U(N)$ and
$\SU(N)$ YM gauge theories with matter on a cylinder and discuss (some
of) its physical consequences. Though the Gribov ambiguity does not
allow a gauge {\em fixing}, we argue that it is possible to maximally
{\em reduce} the gauge freedom.  One is left with an infinite discrete
non-Abelian group $\cG'$ of residual static gauge transformations. It
is possible to determine the (topologically highly non-trivial) space
of all gauge orbits explicitly and to give a general construction of
all physical states which are invariant under $\cG'$. We discuss some
of the implications of this structure for massless QCD(1+1) (i.e.\ YM
theory with massless fermions on a cylinder).  The latter is
especially interesting due to the interplay between the Gribov
ambiguity and the anomaly structure \cite{Jackiw}.

We note that the existence of a non-trivial vacuum structure in
massless QCD(1+1) has already been found in a one-loop calculation by
Hetrick and Hosotani \cite{HH}.  Our construction is without
approximation and applies to all physical states (not only the vacuum).
Moreover, it is a special case of a more general construction valid
for any YM gauge theory on a cylinder.

{}{\bf 2.} The following discussion is in the hamiltonian framework.
The gauge group is $G=\SU(N)$ or $\U(N)$ and assumed in the
fundamental representation.

As mentioned above, a YM-field on a cylinder has a finite number of
{\em physical} degrees of freedom. A simple argument identifying these
is as follows: the temporal component $A_0$ of the YM-field is not
dynamical but only the Lagrange multiplier enforcing Gauss' law (=
invariance under all static gauge transformations), and the only gauge
invariant quantities one can construct at some fixed time from the
spatial component $A_1$ of the YM-field are the eigenvalues of the the
Wilson loop over the whole space\footnote{$P$ is the path-ordering
symbol, $x\in [-\pi,\pi)$ the spatial variable}
\eq
\label{i.2}
W[A_1]=P\exp{\left(-\ii \int_0^{2\pi}\dd{x}A_1(x)\right)} \in G .
\eqend
$W[A_1]$ can always be written as $h^{-1}D h$ with $h\in G$ and $D$ in
the Cartan subgroup of $G$ (note that this representation is not
unique). It is convenient to introduce the basis $\{H_i\}_{i=1}^{N-1}$
in the Cartan subalgebra of the Lie algebra $g$ of $G=\SU(N)$ with
$H_i\equiv e_{ii}-e_{i+1,i+1}$ where $e_{ij}$ is the $N\times N$
matrix with the matrix elements
$(e_{ij})_{kl}=\delta_{ik}\delta_{jl}$. For $G=\U(N)$ we have to add
to these $H_0=1$ (the $N\times N$ unit matrix).  Then we can write $D$
as
\eq
\label{i.2a}
D =\exp{\left(-2\pi\ii\sum_i Y^iH_i \right)}
\eqend
with real $Y^i$. This suggests that the physical YM degrees of freedom
can be represented by these $Y^i$, and that it should be possible to
`fix' the gauge to
\eq
\label{i.3}
A_0(x)=0, \quad
A_1(x)=\sum_{i}Y^iH_i
\eqend
which is essentially the Coulomb gauge $\partial A_1/\partial x=0$.
Indeed, one can systematically eliminate the gauge degrees of freedom
by explicitly solving Gauss' law and end up with the same effective
hamiltonian one obtains by imposing the gauge condition \Ref{i.3}
\cite{LS} (see also Ref.
\cite{HH}).

After that there are still gauge transformations left which act
non-trivially on the $Y^i$ \cite{M,LS}, so \Ref{i.3} is rather a {\em
reducing} than a fixing of {\em the gauge}.  Firstly, there are gauge
transformations permuting the eigenvalues of the diagonalized Wilson
loop $D$: The gauge transformations
\eq
\label{i.4}
p_{ij} = \exp{(\ii \pi
(e_{ij}\ee{\ii\chi} + e_{ji}\ee{-\ii\chi})/4)}
\eqend
($\chi\in\R$ arbitrary) interchange the $i$-th and the $j$-th
eigenvalues and provide a representation $\pi\mapsto \hat{\pi}$ of the
permutation group $S_N$, %\footnote{= group of bijective mappings
%$\pi:\,[N]\equiv\{1,\ldots,N\}\to [N], i\mapsto \pi(i)$}
\eq
\label{i.4a}
Y^i\to (\hat{\pi}Y)^i=\sum_j \hat{\pi}^i_jY^j \quad \forall \pi\in S_N.
\eqend
(Representing the $N$ eigenvalues of $W[A_1]$ as $\ee{\ii Z^i}$ one has
$(\hat{\pi}Z)^i=Z^{\pi^{-1}(i)}$, hence by writing $Y^i=\sum_j
\alpha^i_j Z^j$ and $Z^i=\sum_j \beta^i_j Y^j$ one can easily work out
the matrices $\hat{\pi}^i_j=\sum_k
\alpha^i_{\pi(k)}\beta^k_j$ \cite{examp}).
Secondly, the functions
\eq
\label{i.4b}
h(x)=\exp{(\ii x\sum_j\nu^jH_j)}, \quad \nu^j\in\Z,
\eqend
are gauge transformations {[}i.e.\ obey $h(0)=h(2\pi)${]}. Though
they leave $D$ invariant, they act non-trivially on the $Y^i$:
\eq
\label{i.5}
Y^i\to Y^i +\nu^i, \quad \nu^i\in\Z .
\eqend
Obviously this provides a representation $\nu\mapsto \hat\nu$ of the
Abelian group $\Z^r$ with $r=N-1$ for $G=\SU(N)$ and $r=N$ for
$G=\U(N)$.

It is easy to see that the group $\cG'$ generated by all these gauge
transformations \Ref{i.4a} and \Ref{i.5} is a semidirect product of
$\Z^r$ by $S_N$ \cite{M}.  This residual gauge group gives a full
characterization of the Gribov ambiguity on a cylinder and allows us
to determine the space of gauge orbits explicitly: the set of all YM
fields $A_1(x)$ obeying our gauge condition \Ref{i.3} can be
identified with $\R^r$ ($r$ as above), and $\cG'$ is the group of all
gauge transformations compatable with
\Ref{i.3}. Therefore $\cG'$ generates {\em all} Gribov copies of a
given YM field configuration obeying
\Ref{i.3}, and the space of gauge orbits is $\R^r/\cG'$.
It is interesting to note that (for $G\neq \U(1)$) this is not a
manifold but only an orbifold \cite{Witten}.  From this we can also
easily see that (for $G\neq \U(1)$) there is no gauge fixing on the
cylinder avoiding the Gribov ambiguity: the space $\cA$ of all YM
fields $A_1(x)$ is contractible but $\R^r/\cG'$ is not.  Hence the
fibre bundle $\cA\to \R^r/\cG'$ is non-trivial and does not allow for
a global cross section.

{}{\bf 3.} To dicuss the implications of this in gauge theories with
matter, we refer to the construction of these models mentioned above.
By means of the theory of quasi-free representations of fermion and
boson field algebras \cite{CR,GL} one can construct a Hilbert space
$\cH$ so that the (Fourier components of the) YM field $A_1(x)$, the
matter fields and the matter currents are represented by (closeable)
operators on $\cH$ (with a common, dense, invariant domain \cite{GL}),
the hamiltonian $H$ is a hermitean operator on $\cH$, and there is a
unitary representation $\Gamma$ of the group $\cG$ of all static gauge
transformations on $\cH$ so that $\Gamma(U)H=H\Gamma(U)$ $\forall
U\in\cG$.

The physical Hilbert space of the model is of course
\[\cH_{\rm phys}=\{\Psi\in\cH|\Gamma(U)\Psi=\Psi\quad \forall U\in\cG\},\]
and it can be explicitly constructed in two steps: After imposing the
gauge condition \Ref{i.3}, $\cH$ is reduced to a subspace $\cH_{\rm
phys}'$.  $\cH_{\rm phys}'$ is a tensor product of the Hilbert space
$L^2(\R^r)$ of the YM variables $Y^i$ and a Hilbert space $\cF_M'$ of
the matter fields. On $\cF_M'$ we have a unitary represenation
$\Gamma_M$ of the residual gauge group $\cG'$, and eqs.\ \Ref{i.4a} and
\Ref{i.5} provide a represenation of $\cG'$ on $L^2(\R^r)$.
The subspace $\cH''_{\rm phys}$ of $\cH'_{\rm phys}$ invariant under
the gauge transformations \Ref{i.4b} is obviously spanned by the
states
\eq
\label{i.8a}
\Psi(\vec\theta) = \sum_{\nu_i\in\Z}
\prod_i \exp{(\ii\theta_i(Y^i+\nu^i))}\Gamma_M(h_i)^{\nu^i} \Psi
\eqend
[$\vec\theta\in\R^r$; strictly speaking, we should `smear these out'
by appropriate test functions $f(\vec\theta)$] where $h_i(x)\equiv
\exp{(\ii xH_i)}$ and $\Psi\in\cF_M'$. On $\cH''_{\rm phys}$ we have
then a unitary representation $\hat{\Gamma}$ of the $S_N$ subgroup of
$\cG'$,
\eq
\label{i.8b}
\hat{\Gamma}(\pi)\Psi(\vec\theta) = \Gamma_M(\pi)
\sum_{\nu_i\in\Z} \prod_i \exp{(\ii\theta_i(\hat{\pi}(Y)^i+\nu^i))}
\Gamma_M(h_i)^{\nu^i} \Psi, \quad
\forall \pi\in S_N.
\eqend
Hence the states invariant under the whole residual gauge group $\cG'$,
thus spanning $\cH_{\rm phys}$, are given by
\eq
\label{i.8c}
\f{1}{n!}\sum_{\pi\in S_N} \hat{\Gamma}(\pi)\Psi(\vec\theta),
\quad \Psi\in\cF_M', \vec{\theta}\in\R^r.
\eqend
We can write $\cH_{\rm phys}= S \cH_{\rm phys}''$ where
$S=\f{1}{n!}\sum_{\pi\in S_N} \hat{\Gamma}(\pi)$ is an orthogonal
projection.  It is worth pointing out that for this construction to
work the semidirect product structure of $\cG'$ is crucial, i.e.\ that
$\pi^{-1}\nu\pi \in \Z^r$ for all $\nu \in \Z^r$ and $\pi\in S_N$.

{}{\bf 4.} As an illustration and an especially interesting example,
we discuss in more detail massless QCD(1+1).  Here the very gauge
transformations $h(x)$ \Ref{i.4b} also cause a spectral flow of the
fermions. Due to this, the YM variables $Y^i$ can combine with
fermionic degrees of freedom $Q^i_5$ resulting in variables invariant
under the $h(x)$, and this leads to an interesting additional
structure of the physical states. To show this, we refer to a
(essentially) rigorous construction of massless QCD(1+1) by means the
non-trivial quasi-free representation of the fermion field
operators\footnote{$\sigma,\sigma'\in\{1,2\}$ and $A,B \in \{1,\ldots,
N\}$ are the spin and color indices, respectively}
$\psi^{(*)}(x)\equiv \psi^{(*)}_{\sigma,A}(x)$ naturally associated
with the {\em free} fermion hamiltonian \[
\HH_0=\int_{0}^{2\pi}\dd{x}\normal{\psi^*(x)\gamma_5\f{1}{\ii}
\f{\partial}{\partial x} \psi(x)}, \] with $\gamma_5\equiv
(\gamma_5)_{\sigma\sigma'}=\left(\bma{rr}1&0\\0&-1\ema\right)$
\cite{CR} and $\normal{\cdots}$ normal ordering with respect to the
{\em free} fermion vacuum $\Om$ (see below).  This representation is
on the fermion Fock space $\cF_F$ generated from a state $\Om$ (=
`free Dirac sea') obeying \eq \label{i.6} \hat\psi_{1,A}(n)\Om =
\hat\psi_{2,A}^*(-n-1)\Om =0, \quad \forall A, n\geq 0, \eqend where
$\hat\psi_{\sigma,A}(n) = \int_{0}^{2\pi}\dd{x} \exp{(-\ii
nx)}\psi_{\sigma,A}(x)/\sqrt{2\pi}$ and $n\in\Z$.  After reducing the
gauge to \Ref{i.3}, the remnant of the Gauss' law is $\hat\rho^i(0)
\equiv
\int_0^{2\pi}\dd{x}\normal{\psi(x)H^i\psi(x)}\simeq 0$ \cite{LS} with
the $r$ $N\times N$ matrices $H^i$ obeying $\tra{H^iH_j}=
\del^i_j$.\footnote{i.e.\ $H^i=\sum_j(b^{-1})^{ij}H_j$ with
$(b^{-1})^{ij}$ the inverse matrix to $b_{ij}=\tra{H_iH_j}$} Hence the
reduced Hilbert space $\cH'_{\rm phys}$ of the model is a tensor
product of the space $\cF_F'=\{ \Psi\in\cF_F | \hat\rho^i(0)\Psi = 0,
\quad \forall i\}$ and the Hilbert space $L^2(\R^r)$ for the YM
variables $Y^i$, and there is a unitary representation of the
residual gauge group $\cG'$ on $\cH'_{\rm phys}$.  Especially, the
special gauge transformations $h(x)$ \Ref{i.4b} are implemented by
unitary operators $\Gamma_F(h)$ on $\cF_F'$ \cite{R1,CR}, and they act
non-trivially on the axial `charges' $Q^i_5 \equiv \int_0^{2\pi}\dd{x}
\normal{\psi^*(x)\gamma_5H^i\psi(x)}$ (the $H^i$ as above): \eq
\label{i.7} Q^i_5 \to Q^i_5 + 2 \nu^i \quad \forall i.  \eqend
{[}Proof: We first consider the operators
$Q_{5,AB}=\int_{0}^{2\pi}\dd{x}\normal{\psi^*(x)\gamma_5
e_{AB}\psi(x)}$ (with $e_{AB}$ as above) which can be written as \[
Q_{5,AB} = \sum_{n\geq 0}\left( \hat\psi^*_{1,A}(n)\hat\psi_{1,B}(n)-
\hat\psi_{1,B}(-1-n)\hat\psi^*_{1,A}(-1-n) - \mbox{$\left(\bma{rl}
1\leftrightarrow&\!\!2 \\ n\leftrightarrow&\!\!-1-n\ema
\right)$}\right) .  \] Obviously under the transformation $\psi(x)\to
\exp{(\ii x e_{ii})}\psi(x)$ we have $ \hat\psi^{(*)}_{\sigma,A}(n)\to
\hat\psi^{(*)}_{\sigma,A}(n-\del_{i,A}), $ hence by the canonical
anticommutator relations of the fermion field operators, \[Q_{5,AB}\to
Q_{5,AB} + 2\del_{AB}\del_{iA} = Q_{5,AB} +2(e_{ii})_{AB}.\] From this
we readily deduce that under $\psi(x)\to h(x) \psi(x)$, $Q_{5,AB} \to
Q_{5,AB} + 2\sum_j \nu^j(H_j)_{AB}$, hence as $Q^i_5 =
\sum_{A,B}(H^i)_{BA}Q_{5,AB}$, \[ Q^i_5 \to Q^i_5 +
2\sum_{A,B}(H^i)_{BA}\sum_j \nu^j(H_j)_{AB} \] identical with eq.\
\Ref{i.7}.{]}

{}From this proof it is clear that the implementers $\Gamma_F(h_i)$
create particle-antiparticle pairs in the fermion Fock space (this can
also be seen from the explicit formulas for these derived in
\cite{R1}), hence all physical states $\Psi(\vec\theta)$ \Ref{i.8c}
(including, of course, the groundstate=vacuum of the model) contain a
condensate of fermion-anti-fermion pairs. Moreover, on $\cH_{\rm
phys}$ the YM-degrees of freedom $Y^i$ can be combined with the axial
`charges' $Q^i_5$ to the gauge invariant $\theta$-variables
\eq
\label{i.9}
\theta_i=\f{1}{\ii}
\f{\partial}{\partial Y^i},
\quad p^i\equiv\f{1}{\ii}\f{\partial}{\partial \theta_i}=
Y^i-\f{1}{2}Q^i_5.
\eqend
This can be understood also as a result of the anomalies in this model
\cite{Jackiw}: The naive fermion currents\footnote{the $T^a$ are the
generators of the Lie algebra $g$ of $G$}
$\rho^a(x)=\normal{\psi^*(x)T^a\psi(x)}$ and
$j^a(x)=\normal{\psi^*(x)\gamma_5 T^a\psi(x)}$ are constructed by
means of normal ordering $\normal{\cdots}$ with respect to the {\em
free} fermion vacuum $\Om$. Due to the well-known Schwinger term in
the $[\rho,j]$-commutator
\cite{GO,CR} the currents $j^a(x)$ do not transform covariantly under
static gauge transformations. However, there is a freedom in the
normal ordering prescription \cite{CR} which can be used to fix this
problem: one can show that the (essentially unique) gauge covariant
currents are given by\footnote{we write $\rho(x)\equiv \sum_a
\rho^a(x)T^a$ and similarly for $j$, $A_1$ etc.\ }
$\tilde{\rho}(x)=\rho(x)$ and
$\tilde{\jmath}^a_1(x)=j^a_1(x)-A_1^a(x)/\pi$ (one can think of this as
gauge-covariant normal ordering). The gauge covariant axial `charges'
are therefore $\tilde{Q}_5=\int_0^{2\pi}\dd{x}\tilde{\jmath}(x) = Q_5
- \int_0^{2\pi}\dd{x}A_1(x)/\pi$. By direct calculation one can show that
these give covariantly conserved vector
currents\footnote{$\nu\in\{0,1\}$ is a spacetime index}
$\tilde{\jmath}^\nu=(\tilde{\rho}(x),\tilde
\jmath(x))$:\footnote{$D_\nu$ is the covariant derivative in the
adjoint representation} $D_\nu
\tilde\jmath^\nu(x)=0$, whereas the gauge covariant axial current
$\tilde{\jmath}^\nu_5(x)=(\tilde\jmath(x),\tilde\rho(x))$ has an
anomaly: $D_\nu\tilde{\jmath}^\nu_5(x)=-E(x)/\pi$ (with\footnote{
$\partial_\nu=\partial/\partial x^\nu$, $x^0=t$ (time) and $x^1=x$}
$E=F_{01}= \partial_0 A_1 -\partial_1 A_0 + \ii\ccr{A_0}{A_1}$ the YM
field strength as usual).  In the gauge
\Ref{i.3} this implies $\partial_\nu
\tilde{\jmath}_5(x)= -\sum_i\partial_0 Y^iH_i/\pi$ and gauge invariance
(under the residual gauge transformations in $\cG'$) and
non-conservation of the axial charges $\tilde{Q}_5^i$ identical with
$-2p^i$ ($p^i$ cf.\ eq.\ \Ref{i.9}).

{}{\bf 5.} Usually (especially in path integral approaches) Gribov
ambiguities are accounted for by restricting the YM-field to one
fundamental domain \cite{G,B}. In our case this would correspond to
restricting the $Y^i$ \Ref{i.3} to some appropriate compact subset
$FD$ of $\R^r$ and forgetting about the special gauge transformations
\Ref{i.4a}, \Ref{i.5}. [E.g.\ for $G=\U(N)$, $N\geq 2$, such an
appropriate $FD$ would be given by \[0\leq Z^1\leq Z^2\leq \ldots \leq
Z^N <1\] where $Z^1=Y^0+Y^1$, $Z^i=Y^0+Y^i-Y^{i-1}$ for $i=2,\cdots,
N-1$, and $Z^N=Y^0-Y^{N-1}$.]  Though in principle possible, we think
that (at least on a cylinder) our approach is much more natural.

We finally want to stress that our explicit treatment is restricted in
an essential way to $(1+1)$ dimensions. In higher dimensions
a YM-field has an infinite number of degrees of freedom allowing for a
lot of unitarily inequivalent representations one has to worry about
(only for a finite number of quantum degrees of freedom, all
(reasonable) Hilbert space representation are unitarily equivalent
\cite{BR2}). Moreover, it is only in $(1+1)$ dimensions that the physical
relevant representations of the fermion or boson field algebra
\cite{BR2} of a gauge theory model is unitarily equivalent to the one
of the corresponding free model, and this is crucial for being able to
construct the interacting hamiltonian on the `free' fermion or boson
Fock space, respectively. Nevertheless we think that our results give
an example of a highly non-trivial interplay between the special gauge
transformations implying the Gribov ambiguity and matter fields in
gauge theories, especially when fermions are involved.  It is tempting
to conjecture a similar mechanism also in 4 dimensional QCD which
might be a key to understanding confinement.  We do not know, however,
whether a similar approach to ours can be used there.  As a
prerequisite this would require a {\em full} understanding of the
Gribov ambiguity in 4 dimensions which has not been achieved yet
\cite{B}.

\vspace*{.2cm}
\noindent {\bf Note added:} After submission of this paper we received
Ref.\ \cite{Hetrick} where also the space of gauge orbits for $\SU(N)$
Yang-Mills fields on a cylinder is found.

\begin{center}\subsubsection*{Acknowledgement}\end{center}
One of us (E.L.) would like to thank M. Bergeron, A. Carey, D.
Eliezer, H. Grosse, A. Kovner and M. Salmhofer for helpful
discussions.

\end{document}